\documentclass[sigconf]{acmart}
\topmargin \dimexpr -0.6in

\usepackage[inline]{enumitem}
\usepackage{breqn}
\usepackage{booktabs} 

\copyrightyear{2017} 
\acmYear{2017} 
\setcopyright{acmcopyright}
\acmConference{CIKM'17 }{November 6--10, 2017}{Singapore, Singapore}\acmPrice{15.00}\acmDOI{10.1145/3132847.3133162}
\acmISBN{978-1-4503-4918-5/17/11}

\begin{document}
\title{Soft Seeded SSL Graphs for Unsupervised Semantic Similarity-based Retrieval}

\author{Avikalp Srivastava}
\affiliation{%
  \institution{Indian Institute for Technology, Kharagpur}
}
\email{avikalp22@iitkgp.ac.in}

\author{Madhav Datt}
\affiliation{%
  \institution{Indian Institute of Technology, Kharagpur}
}
\email{madhav@iitkgp.ac.in}

\begin{abstract}
Semantic similarity based retrieval is playing an increasingly important role in many IR systems such as modern web search, question-answering, similar document retrieval etc. Improvements in retrieval of semantically similar content are very significant to applications like Quora, Stack Overflow, Siri etc. We propose a novel unsupervised model for semantic similarity based content retrieval, where we construct semantic flow graphs for each query, and introduce the concept of "soft seeding" in graph based semi-supervised learning (SSL) to convert this into an unsupervised model.

We demonstrate the effectiveness of our model on an equivalent question retrieval problem on the Stack Exchange QA dataset, where our unsupervised approach significantly outperforms the state-of-the-art unsupervised models, and produces comparable results to the best supervised models. Our research provides a method to tackle semantic similarity based retrieval without any training data, and allows seamless extension to different domain QA communities, as well as to other semantic equivalence tasks.



\end{abstract}

\maketitle

\vspace{-0.1in}

\section{Introduction}
Semantic matching and ranking play a key role in various information retrieval and natural language understanding applications such as modern web search, dialogue systems, cross language information retrieval, paraphrasing, textual entailment, question-answering, similar document identification etc. The retrieval model for a query is dependent on the characteristics of the task and can be based on relevance, response/answer to query, entailment, similarity or equivalence \cite{li2014semantic}. In this work we focus on retrieval of semantically equivalent content to a given query. This finds multiple applications in areas of equivalent question retrieval, similar document detection, and paraphrasing applications such as summarization, dialogue, overcoming redundancies in short texts.

With the growing popularity of sites like Stack Overflow, Quora, Baidu Zhidao, and other Q\&A forums, semantic similarity based content retrieval and ranking, particularly in case of questions, is becoming increasingly important, as this enables them to help users see if their questions have already been answered, possibly in a different form, and reduce duplicate content on such websites. 

Detecting semantically similar content is an extremely hard problem mainly because of 3 key factors: 
\begin{enumerate*}
\item the same content can be phrased with very different sentence structure and wording;
\item similarly worded content may represent relation to very different queries; and
\item building training data to capture the diversity and variations in content is very expensive in terms of time and cost.
\end{enumerate*} Thus, naive word-overlap based measures of content similarity such as \cite{broder1997syntactic} do not work for semantic similarity based retrieval problems.

In this paper, we propose a novel unsupervised model for retrieval of semantically equivalent content, where we (a) frame it as a graph-based semi-supervised learning problem; (b) build query induced semantic flow subgraphs for each given query; (c) introduce the concept of "soft seeding" to convert it to an unsupervised solution; and (d) leverage the flexibility of our model by incorporating multiple additional textual features. Our model significantly out-performs the state-of-the-art in unsupervised settings ($38.6\%$ vs $34.1\%$ precision for top-10 retrieval), and produces comparable results to the best supervised models, such as BOW-CNN \cite{dos2015learning}
\vspace{-0.1cm}
\section{Related Work}
Similarity-based retrieval and ranking models are built by developing a similarity function between the query and document objects. Recent works on learning short-text pair similarity focus on representation learning algorithms where the the intrinsic low-dimensional structure in data is exploited to induce similarity. Convolutional neural networks have proven effective in capturing the semantic and syntactic aspects of input sentences, given sufficient amount of training data. \cite{severyn2015learning} uses Pairwise Convolutional Neural Network architecture (PCNN), a supervised model that encodes question-answer pairs into discriminative feature vectors, and produces state-of-the-art performance on question-answering tasks. 

The effectiveness of CNNs in semantic analysis of short texts has motivated their usage in retrieval of equivalent questions. \cite{bogdanova2015detecting} uses CNNs for binary decision making on equivalence, given a question pair, and is extended in \cite{dos2015learning} where the CNN is coupled with a traditional bag-of-words representation to learn hybrid representations for equivalent question retrieval on Stack Exchange Q\&A sites. The CNN and BOW-CNN approaches outperform previous traditional IR approaches like TFIDF \cite{manning2008introduction}, LMDirichlet and LMJelinekMercer \cite{zhai2004study}. However, the primary limitation of such architectures is their dependence on availability of a large amount of high quality training data, making them difficult to apply to different domains of Q\&A datasets, as representations need to be re-learned, and fresh and extensive training data is required. This also considerably inhibits their applicability and extension to other orthogonal applications involving semantic similarity based retrieval. 

The use of stacked context-sensitive autoencoders was introduced by \cite{amiri2016learning}, where the latent representations of a query object and its context are combined into a context-sensitive embeddings. This provides an unsupervised model that produces results comparable to the supervised CNN and BOW-CNN models.
In a supervised setting, this autoencoder model learns weights for combining the context-sensitive representation of question pairs along with additional features derived from BOW representations and word overlap features. The importance of BOW features in question ranking is shown by \cite{dos2015learning} where BOW-CNN significantly outperforms CNN. However, this autoencoder model is limited by its inability to use these additional features in unsupervised settings, where it simply computes the cosine similarity of the context-sensitive representation of the question pair. We leverage our model's flexibility, and exploit properties of SSL graphs by incorporating \textit{feature nodes}, thus allowing introduction of multiple additional features.
Since \cite{amiri2016learning} uses averaged word embeddings as document level vectors, this leads to loss of syntactic and word-level information. Our approach utilizes the recently proposed Word Mover's Distance (WMD) metric \cite{kusner2015word} which elevates high-quality word embeddings to a document metric by formulating the distance between two documents as an optimal transport problem between the embedded words, eliminating the need of obtaining explicit document vector representation. WMD is completely unsupervised and hyper-parameter free, and has achieved unprecedented results on kNN-based document classification and inherently captures the bag-of-words representation of documents.

\section{Soft Seeded SSL Graphs}

Semi-supervised learning approaches have proven successful in scenarios where labeled data is scarce and unlabeled data is present in abundance, and have been rapidly supplanting supervised systems. These approaches leverage unlabeled data by propagating labels via unlabeled samples to capture indirect similarity. This property can specially prove helpful in the problem of semantic equivalence based retrieval in unsupervised setting, where similarity measures that strongly correlate with label assignment are used for discovering indirect similarities through various paths between the data points. Graph-based SSL algorithms based on label propagation are a sub-class of SSL algorithms that use label information associated with initial seed nodes and flow this information in a principled, iterative manner \cite{talukdar2009new} and have been widely used in IR and NLP applications. 

A crucial component of graph-based SSL approaches is the construction of input graph, which should facilitate flow of semantic information among nodes with high pairwise and global feature based similarity measures, for the particular task of propagating semantic equivalence based labels. It is also worth noting that past SSL graph based approaches use a single graph network for global label propagation covering the entire dataset in a transductive setting where small amount of training data is available for assigning the initial seeds. In our case however, we need to work with labels (single-dimension) that correspond to semantic equivalence to the given query, and hence local networks need to be constructed for each query, requiring a new strategy for initial seed assignment.  

We now move on to explain the construction and usage of SSL graphs with respect to the specific task of retrieving top semantically equivalent questions in online Q\&A communities and forums.

\subsection{Query Induced Subgraphs}

Given the question dataset $\mathbf{Q} = \{q_1,q_2,...,q_N\}$, we first seek to learn dense vector representations capturing the underlying semantics for each of the word in the dataset's vocabulary, obtained after pruning of 30 highest frequency words and removal of stopwords, urls and code-tags. We use pre-trained \textit{word2vec} embeddings \cite{mikolov2013distributed} as initialization and further train the model on our dataset \textbf{Q} to account for the presence of specific technical terms and topics in the dataset. \citeauthor{bogdanova2015detecting} \cite{bogdanova2015detecting} demonstrates a significant improvement in performance on equivalent question detection on using in-domain word embeddings as opposed to pre-trained embeddings. We thus obtain an embedding matrix $ X \in {\rm I\!R}^{d \times n}$, where the $i^{th}$ column $\mathbf{x}_i$ represents the d-dimensional embedding for the $i^{th}$ tokenized word.

Given a query question $q_i \in Q$, we obtain a normalized bag-of-words (nBOW) representation $\mathbf{d}_i \in {\rm I\!R}^n $ for $q_i$. To induce a semantic flow subgraph based on this query, we first retrieve \textit{k} questions which share least Word Mover's Distance with $q_i$. To compute WMD between document representations $\mathbf{d}$ and $\mathbf{d}'$, a sparse flow transport matrix $T \in {\rm I\!R}^{n \times n} $ where $T_{i,j} \geq 0$ is computed as a solution of the following linear program:

$$ \min\limits_{\mathbf{T} \geq 0} \sum\limits_{i,j=1}^{n} \mathbf{T}_{ij}c(i,j)$$
subject to: $$ \sum\limits_{j=1}^{n} \mathbf{T}_{ij} = d_i \quad \forall i \in \{1,2,\dots,n\} $$
$$ \sum\limits_{i=1}^{n} \mathbf{T}_{ij} = d'_j \quad \forall j \in \{1,2,\dots,n\} $$

where c(i, j) corresponds to the Euclidean distance between embeddings for word i and word j.

Considering the high computational complexity of WMD metric, it is infeasible to compute the distance of query question with every other question in the dataset, and hence the faster Word Centroid Distance ($||\mathbf{Xd} - \mathbf{Xd'}||_2$) is first used to limit candidates for finding k-nearest neighbors using WMD.  Similar heuristic is applied to each of these \textit{k} '1-hop' neighbors of the query node to obtain a maximum of $k^2$ '2-hop' neighbors under no-overlap, and thus an induced subgraph with maximum $k^2 + k + 1$ nodes. The WMD matrix for the nodes of this subgraph is computed and the distance measures are used as edge weights to obtain a complete subgraph. A choice to drop edges exhibiting distance above a threshold is also available. We thus have a query induced subgraph with semantic flow links, where an important thing to note is that edge weights correspond to node pair distance and not similarity. 

\subsection{Augmenting Feature Nodes}
Traditional document feature representations, such as bag-of-words, IDF-weighted word overlap and other lexical features, significantly improve semantic retrieval when used in conjunction with latent deep learning based representations for texts not limited to single sentences \cite{dos2015learning}. Additionally, one of the weaknesses of approaches relying on word vectors is inability to deal with proper nouns and numbers, which can be found in abundance in general QA forums. Thus the additional features are augmented to the query induced subgraph as \textit{feature nodes}, a method adopted for lexical features in \cite{kannansmart}. Particularly, we use word overlap features, common n-grams and skip-grams (up to length 3) as feature nodes ($V_F$), in addition to nodes ($V_R$) in the query induced subgraph. An edge ($v_f, v_r$) for $v_f \in V_F$ and $v_r \in V_R$ with weight as IDF value of $v_f$ is added if $v_f$ belongs to feature set of $v_r$. 
We would also like to point out that BOW representation of documents is inherently considered in the computation of their Word Mover's Distance.

\subsection{Soft Seeding}
Being a subset of semi-supervised learning based approaches, SSL graphs naturally require a small amount of initial training labels, which are used as seed labels for the seed nodes for providing the information that propagates through the graph network. Even in presence of some supervised data, this is difficult to incorporate in SSL graphs that are constructed as query induced subgraphs, since some seed labels signaling equivalence/non-equivalence will be required for each query. 

Here, we describe our method of introducing soft seeds in query induced SSL graphs, which serves the dual purpose of combating the problem of unavailability of initial seeds and incorporating multiple measures of semantic similarity in the framework. Our main objective is to be able to choose nodes in the induced subgraph which can be assigned as 1-seeds (i.e. with initial soft label = 1.0 corresponding to high semantic equivalence with query node) and 0-seeds (low semantic equivalence with query node), and later adapt the SSL graph objective function to take into account the softness of these seeds, as compared to the hard training data based seeds available otherwise. One hard seed available is the query node (with label = 1.0) which acts as a source of flow of high semantic equivalence to it's neighboring nodes weighted by the edge weights.

Assignment of the initial soft seeds requires a similarity measure different from the one used to construct the graph and feature links to avoid conditioning. An important semantic similarity measure exploited in question retrieval tasks is topic modeling based document representation. \cite{amiri2016learning} uses sparse non-negative matrix factorization (NMF) to obtain topic and context matrices, and feed the context vectors to its autoencoder architecture as the context of the query question. In the next section, we explain our choices for the topic modeling approaches for obtaining 1-seeds and 0-seeds followed by the optimization objective for our final query induced SSL graph. 

\textbf{(I) Thresholded SVD based Topic Model for 1-seeds}

NMF exhibits a natural clustering property, however our approach towards assigning 1-seeds with high confidence requires not only analyzing the questions/documents in the same topic induced cluster as the query document, but also requires that this topic forms the dominant topic for the query and that within this cluster a further refined score among the documents can be obtained for assigning more accurate near-to-1 soft labels. Recent work by \cite{bansal2014provable} gives a SVD-based algorithm followed by thresholding on the data matrix that provides bound $l_1$ reconstruction error under the simplifying and empirically supported assumptions of documents being drawn from \textit{dominant admixtures} and presence of topic specific \textit{catchwords}. We briefly explain the assumptions with the help of parameters $w_0$ (lowest probability that a topic is dominant), and non-negative reals $\alpha, \beta, \rho, \delta, \epsilon, \epsilon_0$ satisfying: $\, \beta + \rho \leq (1-\delta)\alpha, \quad \alpha + 2\delta \leq 0.5, \quad \textrm{and} \quad \delta \leq 0.08$

(a) \textbf{Dominant Admixture assumption}: (i) If for every document, there is one topic whose weight is significantly higher than the other topics i.e. for a document $j$ with dominant topic $l(j)$, $W_{l(j),j} \geq \alpha$ and $ W_{l'j} \leq \beta, \quad \forall l' \neq l(j)$;
(ii) there exists at-least a small fraction ($\epsilon_0 w_0 s$) of documents which are almost purely ($\geq 1 - \delta$) on that topic.

(b) \textbf{Catchwords assumption}: Presence of a group of words, say $\left\lbrace S_1, S_2, \dots ,S_k \right\rbrace$, for each of the $k$ topics which together occur with high probability ($\sum_{i \in S_l} M_{il} \geq p_0$) and that each individual word in the group occurs more frequently in that topic than any other ($\forall i \in S_l, \forall l' \neq l, M_{il'} \leq \rho M_{il} $)

We utilize these simplifying assumptions to learn topics from dominant admixtures for the given dataset \textbf{Q}. Sparse thresholding is applied to the data matrix $\mathbf{A}$,  where threshold for the $i^{th}$ word is determined as follows (given \textit{m} is average number of words per document):
Let $\zeta_i$ be the highest value of $\zeta \in \{0,1,\dots,m\}$ such that $|\{j:A_{ij} > \frac{\zeta}{m}\}| \geq \frac{w_0 s}{2}$; $ |\{j:A_{ij} = \frac{\zeta}{m}| \leq 3\epsilon w_0 s$. The thresholds ($\zeta$) are used to obtain the thresholded matrix \textbf{B}:\begin{equation*}
    B_{ij}=
    \begin{cases}
      \sqrt[]{\zeta_i}, & \text{if}\ A_{ij} > \frac{\zeta_i}{m} \\
      0, & \text{otherwise}
    \end{cases}
  \end{equation*}

The singular values of this thresholded matrix \textbf{B} are provably bounded, and this condition is used by the authors to prove that the clusters ($R_1,R_2,\dots,R_k$) generated by Lloyd's k-means clustering on the columns of \textbf{B} with initial starting centers as the k-means clustering centers of columns of SVD-based k-rank approximation $\mathbf{B}^{(k)}$ of \textbf{B} correctly identify almost all the documents' dominant topics. Thus all the documents belonging to the same cluster as the query document are candidates for 1-seeds. We now obtain a refined score based on the sum of document terms over identified catchwords. We bypass the posterior inference of document topic distribution, and compute the sparse catchword-based topic weights for each document. For the set of catchwords $J_l$ for topic l, the weight of topic l in document j is calculated as $\sum_{i \in J_l} A_{ij} $ which evaluates to zero for majority of topics giving us a sparse catchword based representation, whose cosine similarity with query representation is finally utilized to retrieve top k' seeds and assign the similarity value as soft label.

\textbf{(II) Thresholded SVD for 0-seeds}

To assign 0-seeds with high confidence, we analyze the topic-word distribution vectors in the topic matrix obtained by thresholded SVD algorithm. The vector $M_{.,l'}$ with maximum euclidean distance from $M_{.,l}$, where l corresponds to the dominant topic for query document is selected. For each document in the \textit{l'} cluster, we analyze the topic differential score \cite{bairi2016framework} for topic \textit{l} with reference to the query document, and select the k' documents with lowest topic differential scores as the 0-seeds with their topic differential scores as soft labels.

\subsection{Optimization Objective}
The optimization objective for our final SSL varies from the standard objective in 3 aspects: (a) The label dimension for each node is simply 1, (b) the edge weights linking the primary nodes of the graph signify distance and not similarity and (c) we account for the softness of the initial seeds by relaxing the penalty on change of seed label values from their initial values, and introduce dropout on the edges linked to the soft seeds with probability $p$. A localized form of over-fitting occurs, where nodes adjacent to soft-seeded nodes, learn much of their value from those nodes, and may end up having much greater fit/agreement with the localized sub-graph structure instead of the overall graph. We leverage its analogous nature to training a neural network, and counter this effect by reducing the influence of the local sub-graph via dropout \cite{srivastava2014dropout}. 

\begin{dmath*} 
s_i(\hat{C}_i - C_i)^2 + \mu_{pp}(\hat{C}_i - U)^2 + \mu_{np}(\sum\limits_{j \in N_F(i)}w_{ij}(\hat{C}_i - \hat{C}_j)^2 + \sum\limits_{k \in N_R(i)}w_{ik}(\hat{C}_i - \hat{C}_k)^2)
\end{dmath*}

where $\hat{p}*s_i$ is an indicator for seeds that corresponds to $max(s, 1-s)$ for their initial seed value $s$, and is $0$ for non-seeds and $\hat{p}$ is $1$ in case of no dropout and $0$ otherwise. $\hat{C}_i$ denotes the learned current label value of node \textit{i}, $C_i$ is the initial assigned values for seeds. U corresponds to the uniform prior (0.5), and $N_R(i)$ and $N_F(i)$ denote the graph node and feature node neighbor set for $i^th$ node. Since $w_{ij}$ are WMD based distances, we modify this loss function to penalize label similarity rather than separation. Thus the term in the function is changed to  $(\sum\limits_{j \in N_F(i)}w_{ij}(1 - (\hat{C}_i - \hat{C}_j)^2))$, where the constant term is the sum of weights of all outgoing edges from node \textit{i} and hence can be dropped. Thus, the only change is that this term is subtracted rather than added in the loss function. We follow the vanilla Jacobi iterative update and set the termination conditions threshold on maximum change in a label's distribution \cite{ravi2016large}  

\section{Experimental Evaluation}
We evaluate our model on the "\textit{qSim}" dataset with a question ranking task, where given a test question,
the model should retrieve the top $k$ questions that are semantically equivalent to the test question for $k \in \mathcal{K},~ \mathcal{K}= \left\lbrace 1, 5, 10 \right\rbrace$. \textit{qSim} is a community question-answer dataset scraped from Stack Exchange along with ground-truth labels for semantically equivalent questions from \cite{dos2015learning}. For the sake of direct comparison, we use the same test data distribution given in \cite{dos2015learning} and used by \cite{amiri2016learning}. Statistics of the dataset are shown in Table 1.
The parameter settings adopted for conducting these experiment involve k = 50, 300 and 750 nearest neighbor retrievals for the query induced subgraph for Precision @1, @5 and @10 tasks respectively, where WCD measure was used initially to retrieve 500, 3000 and 7500 documents respectively. For application of TSVD,  we used number of topics as k = 200, parameters $ w_0 = \frac{1}{k}, \epsilon = \frac{1}{10}, \epsilon_0 = \frac{1}{6} $.
In the following section, our proposed model, Soft Seeded SSL Graph, is abbreviated as \textbf{SSG}. We use \textbf{SSG-D} and \textbf{SSG-WD} to refer to our model with dropout and without dropout respectively.
\vspace{-0.2cm}
\begin{table}[!htbp]
\centering
\caption{\textit{qSim Dataset Statistics}}
\label{my-label}
\begin{tabular}{|c|c|c|}
\hline
\textbf{Split} & \textbf{Pairs} & \textbf{\%Positive Pairs} \\ \hline \hline
Train          & 205K           & 0.048\%                   \\ \hline
Dev            & 43M            & 0.001\%                   \\ \hline
Test           & 82M            & 0.001\%                   \\ \hline
\end{tabular}
\end{table}
\vspace{-0.2cm}
We compare our unsupervised Soft Seeded SSL Graph based model against the supervised PCNN and PBOW-PCNN models presented in \cite{dos2015learning}, the supervised SAE and CSAE models, and the unsupervised SAE and CSAE models as presented in \cite{amiri2016learning}. We also compare our model against the supervised SAE-DST and CSAE-DST models which include additional element-wise similarity features as described in \cite{amiri2016learning}. We use \textit{Precision at Rank $k$}, denoted by $P@k, ~ k \in \mathcal{K}$, to evaluate performance of our model. Results in table 2 demonstrate the effects of dropout in our model.
\vspace{-0.2cm}
\begin{table}[!htbp]
\centering
\caption{Question ranking precision for our model, without and with dropout respectively}
\label{my-label}
\begin{tabular}{|c|c|c|c|}
\hline
\textbf{Model} & \textbf{P@1}  & \textbf{P@5}  & \textbf{P@10} \\ \hline
SSG-WD         &  \textbf{19.0 }            & 34.6              &    38.1           \\ \hline
\textbf{SSG-D} & 18.9 & \textbf{34.9} & \textbf{38.6} \\ \hline
\end{tabular}
\end{table}
\vspace{-0.2cm}
\begin{table}[!htbp]
\centering
\caption{Question ranking precision in \textit{supervised} settings}
\label{my-label}
\begin{tabular}{|c||c|c|c|}
\hline
\textbf{Model} & \textbf{P@1}  & \textbf{P@5}  & \textbf{P@10} \\
\hline
\hline
PCNN           & 20.0          & 33.8          & \textbf{40.1} \\
\hline
SAE            & 16.8          & 29.4          & 32.8          \\
\hline
CSAE           & \textbf{21.4} & \textbf{34.9} & 37.2          \\
\hline
\hline
PBOW-PCNN      & 22.3          & \textbf{39.7} & \textbf{46.4} \\
\hline
SAE-DST        & 22.2          & 35.9          & 42.0          \\
\hline
CSAE-DST       & \textbf{24.6} & 37.9          & 38.9         \\
\hline
\end{tabular}
\end{table}

\section{Conclusion}
As can be observed from Tables 3 and 4, our completely unsupervised model for semantic similarity based content retrieval, significantly outperforms the state-of-the-art unsupervised models, by $0.3\%, 1.7\%$ and $4.5\%$ for question ranking precision for top-1, 5 and 10 retrieval, respectively. It also produces results comparable to the best supervised models, in higher ranks (top-5, 10 retrieval). In this paper, we also proposed a new domain of application of SSL graphs for query based retrieval, and subsequently introduced \textit{soft seeding} as a strategy to make our model completely unsupervised, thus, enabling extension to different domain QA communities, and to other semantic equivalence tasks.

\begin{table}[!htbp]
\centering
\caption{Question ranking precision for models in \textit{unsupervised} settings, including our model, with dropout}
\label{my-label}
\begin{tabular}{|c||c|c|c|}
\hline
\textbf{Model} & \textbf{P@1}  & \textbf{P@5}  & \textbf{P@10} \\ \hline \hline
SAE            & 17.3          & 32.4          & 32.8          \\ \hline
CSAE           & 18.6          & 33.2          & 34.1          \\ \hline
\textbf{SSG-D} & \textbf{18.9} & \textbf{34.9} & \textbf{38.6} \\ \hline
\end{tabular}
\end{table}

\section{Future Work}
While our paper is focused on proposing a novel model for semantic similarity based content retrieval, it also provides an adaptive framework for solving any problem that could be framed as a graph-based semi-supervised learning task, in an unsupervised manner. This improves possibilities for research and applications of semantic matching and ranking in domains where labeled data in unavailable. We are currently extending this work for faceted recommendation of web news articles, and preliminary results are encouraging.


\bibliographystyle{ACM-Reference-Format}
\small{
\bibliography{sigproc} 
}

\end{document}